\documentclass[journal]{IEEEtran}
\usepackage{amsmath,amsfonts}
\usepackage{algorithmic}
\usepackage{array}
\usepackage[caption=false,font=normalsize,labelfont=sf,textfont=sf]{subfig}
\usepackage{textcomp}
\usepackage{stfloats}
\usepackage{url}
\usepackage{verbatim}
\usepackage{graphicx}
\usepackage{enumitem}
\def\BibTeX{{\rm B\kern-.05em{\sc i\kern-.025em b}\kern-.08em
    T\kern-.1667em\lower.7ex\hbox{E}\kern-.125emX}}
\usepackage{balance}
\usepackage{booktabs}
\usepackage{multirow}
\usepackage{algorithm}
\usepackage{xcolor}
\usepackage{amsmath}
\newcommand{\OUTPUT}{\item[\textbf{Output:}]}

\begin{document}
\title{Synthetic Abundance Maps for Unsupervised Super-Resolution of Hyperspectral Remote Sensing Images}

\author{Xinxin Xu, Yann Gousseau, Christophe Kervazo, Saïd Ladjal
\thanks{The authors are with LTCI, Télécom Paris, Institut Polytechnique de Paris, 91120 Palaiseau, France (e-mail: xinxin.xu@telecom-paris.fr; yann.gousseau@telecom-paris.fr; christophe.kervazo@telecom-paris.fr; said.ladjal@telecom-paris.fr).}
}

\markboth{}
{Synthetic Abundance Maps for Unsupervised Super-Resolution of Hyperspectral Remote Sensing Images}

\maketitle

\begin{abstract}
Hyperspectral single image super-resolution (HS-SISR) aims to enhance the spatial resolution of hyperspectral images to fully exploit their 
spectral information. While considerable progress has been made in this field, most existing methods are supervised and require ground truth data for training—data that is often unavailable in practice. To overcome this limitation, we propose a novel unsupervised training framework for HS-SISR, based on synthetic abundance data, where no high-resolution ground-truth reference is required for training. The approach begins by unmixing the hyperspectral image into endmembers and abundances. A neural network is then trained to perform abundance super-resolution using synthetic abundances only. These synthetic abundance maps are generated from a dead leaves model whose characteristics are inherited from the low-resolution image to be super-resolved and from the known point spread function (PSF) of the hyperspectral sensor. This trained network is subsequently used to enhance the spatial resolution of the original image’s abundances, and the final super-resolution hyperspectral image is reconstructed by combining them with the endmembers. Experimental results demonstrate both the training value of the synthetic data and the effectiveness of the proposed method across 3 datasets, 3 scaling factors, and several evaluation metrics. The code is available at
\url{https://github.com/xinxinxu99/SISR-DL.git}
\end{abstract}


\section{Introduction and Related Work}
\IEEEPARstart {H}{yperspectral} images (HSI) are three-dimensional data structures, similar to multispectral and natural images, composed of two spatial dimensions and one spectral dimension. While natural images are typically acquired in three RGB bands and multispectral images in a few dozen bands, HSIs capture scenes across hundreds or even thousands of contiguous spectral bands, depending on the sensor. This high spectral resolution allows for fine-grained material discrimination and supports a wide range of applications, such as astrophysics \cite{fahes2021unrolling}, target detection \cite{wu2022uiu}, and numerous remote sensing tasks, which are of primary interest in the context of this work. In particular, HSIs have proven highly effective for vegetation monitoring \cite{mills2010evaluation}, land use and land cover classification \cite{yao2023extended}, to name a few.

Regardless of the application, hyperspectral data generally suffer from low spatial resolution due to the need to capture narrow spectral bands, which limits the amount of photons per pixel. To preserve a sufficient signal-to-noise ratio, sensors use larger pixels, which reduces spatial precision. This limitation is particularly critical in remote sensing, where the distance between the sensor and the scene further degrades spatial precision. To address this challenge, super-resolution techniques aim to improve the spatial resolution of HSIs while preserving their spectral consistency. Existing methods for hyperspectral super-resolution can be grouped into two main categories:
\begin{list}{}{\setlength{\leftmargin}{0pt}}

    \item \textbf{MSI-HSI fusion} : This strategy seeks to combine a high-spatial, low-spectral resolution multispectral image (MSI) with a low-spatial, high-spectral resolution HSI. A widely used intermediary step for this purpose is hyperspectral unmixing \cite{yokoya2017hyperspectral}, which decomposes each pixel into a linear combination of endmembers (pure material spectral signatures) and their corresponding abundances \cite{bioucas2012hyperspectral}. The key idea is to extract the spatial details from the MSI (via its abundance) and to combine them with the spectral information provided by the endmembers of the HSI. Since the MSI provides a high-resolution spatial reference, most fusion-based approaches do not require access to a ground-truth super-resolved HSI during training; instead, they can be trained in an unsupervised manner by enforcing consistency with the observed MSI and HSI through the degradation model \cite{aburaed2023review}. Several deep learning-based methods have extended this principle. For example, HyCoNet \cite{zheng2020coupled} introduces a three-branch autoencoder structure capable of jointly estimating the point spread function (PSF) and spectral response function (SRF) from the data. More recently, Hong et al. \cite{hong2023decoupled} proposed a fusion architecture that decouples the MSI and HSI into common and sensor-specific components before recombining them in a shared feature space to produce a fused output image.

    \item \textbf{Single Image Super-Resolution (SISR)} :In contrast to fusion methods, SISR approaches aim to reconstruct a high-spatial resolution HSI from a single low-spatial resolution observation, without relying on external guidance from an MSI. Early SISR techniques were model-driven. For instance, Akgun et al. \cite{akgun2005super} formulated super-resolution as an inverse problem, solved using Projection Onto Convex Sets (POCS) \cite{bauschke1996projection}. With the rise of deep learning, data-driven methods have become predominant. A common approach is to use Convolutional Neural Networks (CNNs) that can learn the mapping from low- to high-resolution directly from training data. Yuan et al. \cite{yuan2017hyperspectral} leveraged transfer learning by training networks on natural images and adapting them to hyperspectral data.

    To better capture the intrinsic spatial-spectral correlations in HSIs, several network architectures have been designed to jointly process both domains. Early deep models either used fully 3D convolutions \cite{mei2017hyperspectral} to handle spectral cubes holistically, or decoupled spatial and spectral extraction via 1D (spectral) and 2D (spatial) convolutions \cite{li2019dual}. More advanced designs propose hybrid schemes to balance performance and efficiency. For example, SSPSR \cite{jiang2020learning} introduces a spatial-spectral residual block combining 2D convolutions and spectral attention, while adopting spectral grouping to reduce parameter overhead. HSISR \cite{li2022hyperspectral} builds upon this by incorporating a parallel RGB super-resolution task, enabling shared spatial feature learning across modalities and semi-supervised training on unlabelled HSI data. MCNet \cite{li2020mixed} further develops this idea with a mixed 2D/3D convolutional architecture, where spatial features are extracted via lightweight 2D convolutions and spectral dependencies are modeled through separable 3D convolutions within a unified residual framework. More recently, transformer-based architectures have been introduced for HSI super-resolution. MSDformer \cite{chen2023msdformer} employs a multiscale deformable transformer that alternates spatial–spectral attention and deformable sampling to capture long-range spatial–spectral dependencies while adapting to complex structures. At a larger scale, HyperSIGMA \cite{wang2025hypersigma} is a hyperspectral foundation model based on a sparse sampling attention mechanism and spectral enhancement modules; although designed as a general-purpose representation learner, it can be fine-tuned for SISR and has demonstrated strong performance on this task thanks to its large-scale pre-training.
\end{list}

Despite their potential, both fusion and SISR methods face critical limitations. Fusion techniques require precise spatial alignment between the MSI and HSI, which is difficult to achieve in practice and can introduce artifacts or performance degradation when misalignments occur. This issue is particularly acute in remote sensing, where, to the best of our knowledge, co-registered MSI and HSI pairs are not available. On the other hand, most SISR methods rely on supervised learning, which demands large datasets with known high-resolution ground truth. However, such datasets are rare in hyperspectral remote sensing due to acquisition constraints, making it challenging to train and evaluate these models.

A possible way to overcome the scarcity of data is to rely on synthetic data for training models. Generating such data in the field of hyperspectral imaging has been little explored and remains a challenging task. Audebert et al. \cite{audebert2018generative} demonstrated that GANs can be used to generate realistic, class-conditioned hyperspectral samples, despite known issues like overfitting and training instability. Liu et al.\cite{liu2023diverse} introduced a diffusion-based method to create synthetic HSIs from RGB images and the USGS Spectral Library V7 \cite{kokaly2017usgs}, with a particular focus on capturing spectral variability in the resulting synthetic images. However, their approach remains limited by the availability and completeness of spectral libraries.

In the field of computer vision, the use of synthetic data is much more common, and many works have been devoted to this topic, mostly for classification, estimation or detection tasks, relying for instance on simplified object generation \cite{dosovitskiy2015flownet} or more recently text-to-image models \cite{tian2023stablerep}. For image restoration tasks, it was recently proposed to rely on a simple occlusion-based geometric model, the dead leaves model \cite{matheron1968modele, alvarez1999size}, to perform fully synthetic training of restoration models \cite{achddou2023fully}. Dead leaves models offer a simple framework for simulating edges and homogeneous regions by overlapping opaque objects and effectively reproduce non-Gaussian statistics observed in natural images. Experiments performed in \cite{achddou2023fully} show that convolutional neural networks trained exclusively on dead leaves images can achieve competitive performance for the denoising and  super-resolution of natural images.

Inspired by these advances, we propose to adapt this concept to hyperspectral remote sensing images with a particular attention to  urban scenes. In this work, we introduce a new framework for generating synthetic abundances images based on the dead leaves model, tailored to the characteristics of low-resolution remote sensing data. This approach enables the training of super-resolution neural networks without relying on any real high-resolution hyperspectral ground truth data, thus removing a major bottleneck in supervised learning. The contributions of this study are threefold:

\begin{enumerate} 

    \item We demonstrate the feasibility of training super-resolution networks for hyperspectral images entirely on synthetic data. This strategy leverages the known PSF of the hyperspectral sensor to ensure realistic data generation and enables unsupervised learning, as no paired real ground-truth data are required, thereby overcoming the common limitation of scarce or unavailable references.

    \item We introduce a synthetic HS data generation process designed specifically for remote sensing scenarios, and show that it leads to competitive super-resolution results.

    \item We propose to generate synthetic HSI through hyperspectral unmixing, which serves as a form of spectral compression. This approach is both lightweight and computationally efficient, significantly reducing the cost and complexity of data synthesis.
    
\end{enumerate}

This article is organized as follows: Section \ref{sec_proposed_method} describes the proposed method and Section \ref{sec_Experiments} presents the ablation studies,  the experimental results and discussion.

\section{Proposed method}
\label{sec_proposed_method}

\subsection{Approach Overview}

\begin{figure*}
    \centering
    \includegraphics[width=1\linewidth]{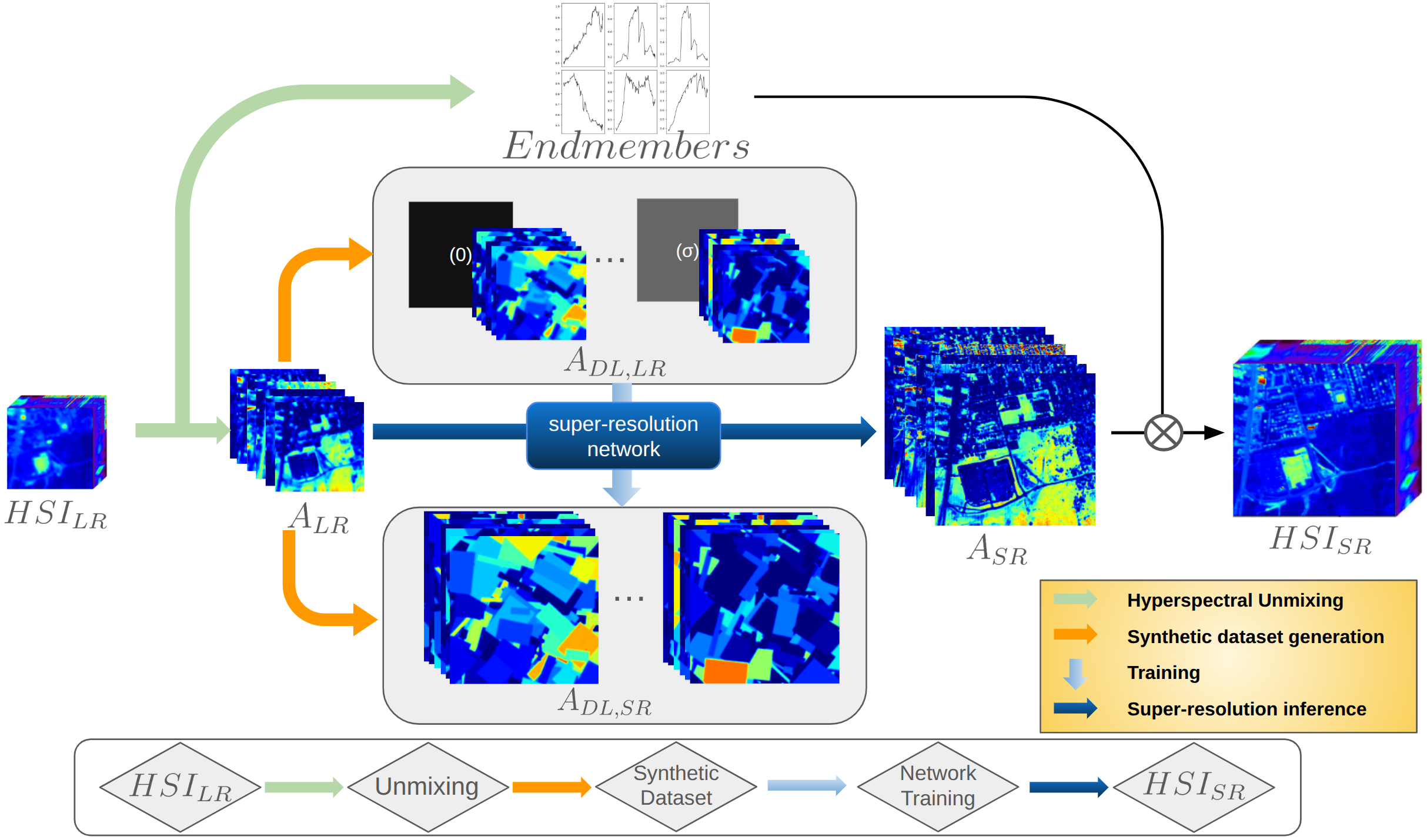}
    \caption{Structure of the proposed method: The super-resolution network is trained with synthetic pairs $(A_{DL,SR},A_{DL,LR})$ and then used to super-resolve $A_{LR}$ into $A_{SR}$.}
    \label{Fig:Structure}
\end{figure*}

Super-resolution consists in obtaining a high spatial resolution hyperspectral image $HSI_{HR} \in \mathbb{R}^{H \times W \times L}$ from a low-resolution image $HSI_{LR} \in \mathbb{R}^{h \times w \times L}$, where $h$, $w$, $H$, and $W$ (with $h < H$ and $w < W$) denote the spatial dimensions, and $L$ is the spectral dimension. Despite its simplicity and associated limitations \cite{kervazo2021provably}, we assume that each pixel in the HSI can be decomposed using the following linear hyperspectral unmixing model:

\begin{equation}
HSI_{LR}(i,j,l) = \sum_{m=1}^{M} A_{LR}(i,j,m)  \cdot S(m,l) + N_{LR}(i,j,l)
\label{eq:unmixing}
\end{equation}

\noindent where $A_{LR} \in \mathbb{R}^{h \times w \times M}$ and $S \in \mathbb{R}^{M \times L}$ represent the abundances and the endmembers of the low-resolution HSI, respectively, $M$ the number of materials and $N_{LR}\in \mathbb{R}^{h \times w \times L}$ is an additive Gaussian white noise. In practice, we apply the Minimum Volume (MinVol) hyperspectral unmixing method on $HSI_{LR}$ \cite{gillis2014successive} to extract the endmembers $S$, using a minimum‑volume NMF with $M = 6$. This method solves a nonnegative matrix factorization problem with a log‑determinant volume regularization term, enforcing sum‑to‑one constraints on the abundances and automatically tuning the regularization parameter to balance data fidelity and volume minimization. The initialization is given by the Successive Nonnegative Projection Algorithm (SNPA), followed by alternating projected fast‑gradient updates of the endmembers and abundances, and the iterations are stopped after 1000 iterations when the relative variation of the objective function becomes negligible. Then, to limit reconstruction errors and to simplify the calculation of correlated noise described below, the abundances $A_{LR}$ are computed using the least squares method.

Based on this decomposition, the main idea of our method is to generate a large number of synthetic abundances $A_{DL,LR} \in \mathbb{R}^{h \times w \times M}$ and $A_{DL,HR} \in \mathbb{R}^{H \times W \times M}$ from $A_{LR}$, using the dead leaves model (DL for short). The high-resolution abundances $A_{DL,HR}$ are first generated using the DL model, while the corresponding low-resolution abundances $A_{DL,LR}$ are obtained by applying a PSF simulated through Gaussian blurring and bicubic downsampling. To improve the robustness of the network training process, a low correlated noise $N_{A, LR}\in \mathbb{R}^{h \times w \times M}$  is added to a randomly selected subset of the low-resolution abundances $A_{DL,LR}$. The synthetic pairs $(A_{DL,HR}, A_{DL,LR})$ are then used to train a super-resolution neural network. During inference, the abundances $A_{LR}$ of the HSI to be super-resolved are fed into the network to estimate the super-resolved abundances $A_{SR}$. The final super-resolution hyperspectral image $HSI_{SR}$ is reconstructed using $HSI_{SR}(i,j,l) = \sum_{m=1}^{M} A_{SR}(i,j,m) \cdot S(m,l)$. The structure of our method is summarized in Fig.~\ref{Fig:Structure}.

The following sections describe in detail the process of generating synthetic abundances, the super-resolution network and the addition of noise.

\subsection{Synthetic Abundance Generation by the Dead Leaves model}

\begin{figure}
    \centering
    \includegraphics[width=1\linewidth]{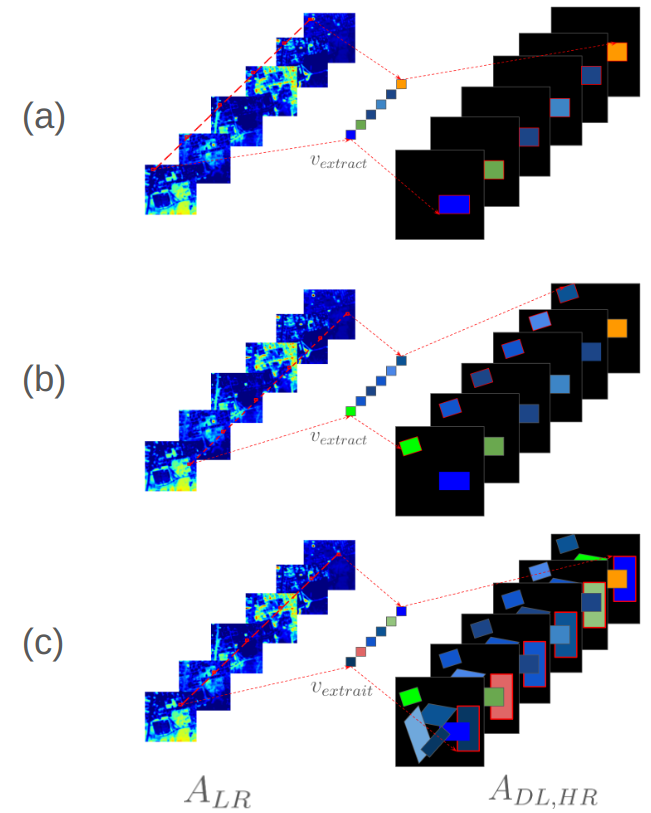}
    \caption{
        Illustration of synthetic abundance generation for $A_{DL,HR}$ using the $1^{\text{st}}$, $2^{\text{nd}}$, and $3^{\text{rd}}$ value columns of $v_{\text{extract}}$ in (a), (b), and (c), respectively. 
        In (c), an example highlights the occlusion mechanism: a newly added leaf is partially covered by a previously generated one, illustrating the stacking behavior of the dead leaves model.
    }
    \label{fig_DL_exemple}
\end{figure}

\begin{figure}
    \centering
    \includegraphics[width=1\linewidth]{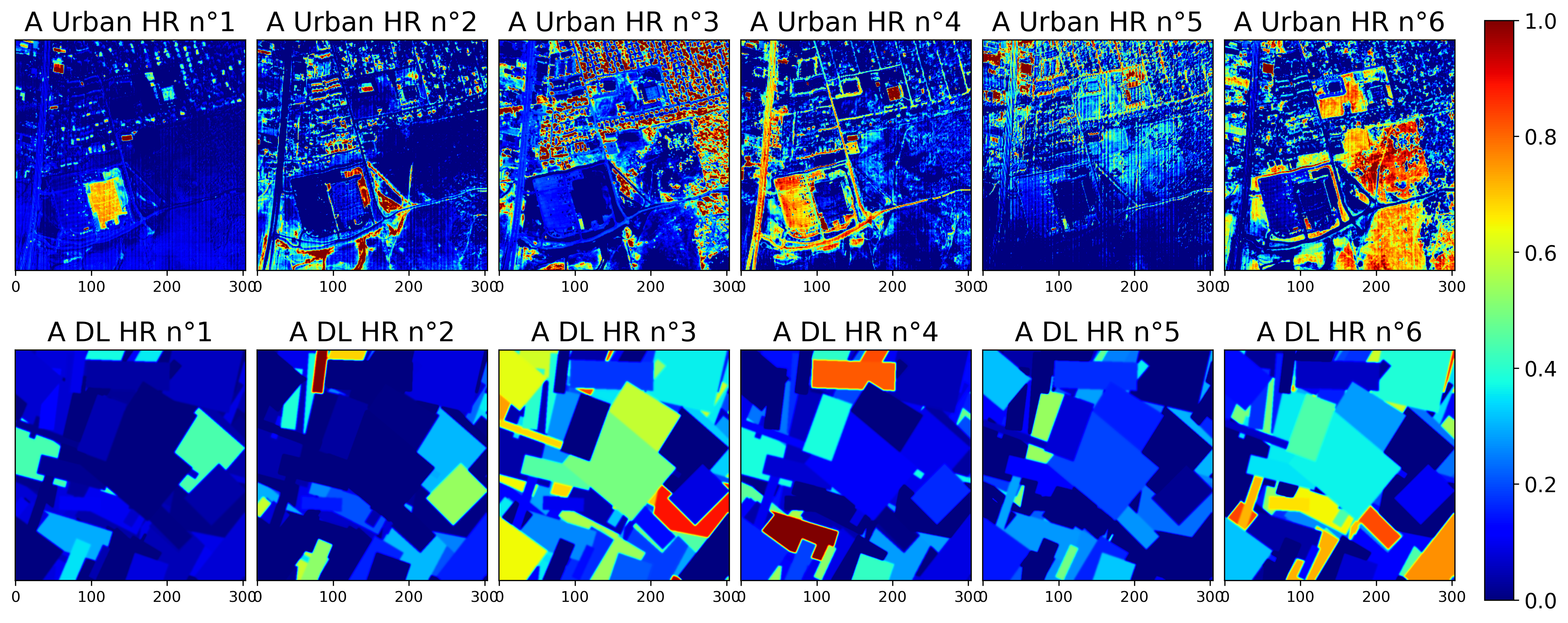}
    \caption{Comparison between real urban abundance (Top) and synthetic dead leaves abundance (Bottom).}
    \label{fig_compare_DL}
\end{figure}

\begin{figure}
    \centering
    \includegraphics[width=1\linewidth]{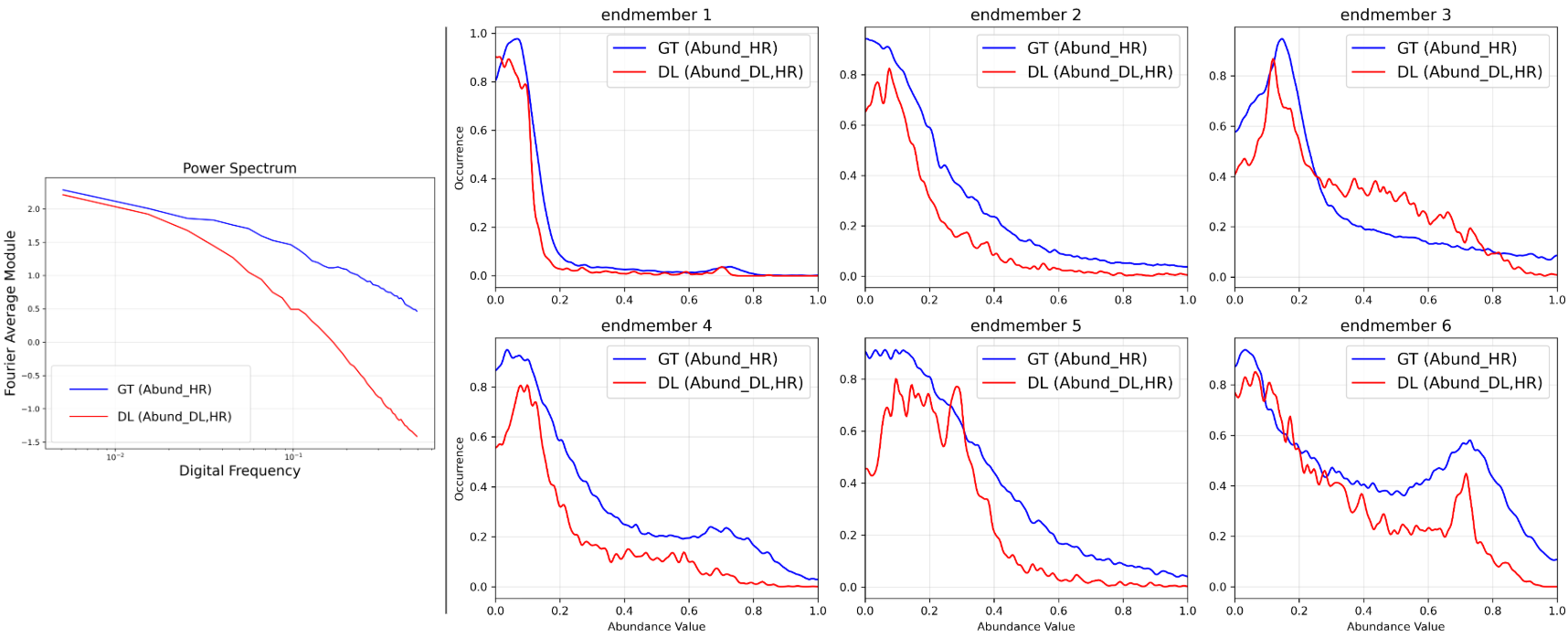}
    \caption{Power spectrum (Left) and per-endmember abundance histograms (Right) between real urban abundance (Blue) and synthetic dead-leaves (Red).}
    \label{fig_PowSpec_Hist}
\end{figure}

\begin{algorithm} 
\caption{Synthetic Abundance Generation for a Single LR-HR Pair with the Dead Leaves Model}
\label{alg:dead_leaves}
\begin{algorithmic}[1]
\REQUIRE Real abundance $A_{LR} \in \mathbb{R}^{h \times w \times M}$, target resolution $(H, W)$  
\OUTPUT Synthetic high-resolution abundances $A_{DL,HR}$ and low-resolution abundances $A_{DL,LR}$  

\STATE Initialize $A_{DL,HR} \gets 0 \in \mathbb{R}^{H \times W \times M}$
\STATE Initialize $Mask \gets \emptyset$
\WHILE{$\# Mask < HW$}
    \STATE Sample $Leaf \gets \text{Rect}(a, b, \theta) + (x,y)$  
    \STATE Extract a material proportion vector $\textbf{v}_{extract} \gets \text{RandomPixel}(A_{LR})\in\mathbb{R}^M$ 
    
    \FOR{$(x,y) \in Leaf \backslash Mask$}
        \STATE $A_{DL,HR}[x,y,.] \gets \textbf{v}_{extract}[.]$
    \ENDFOR
    \STATE $Mask \gets Mask \cup Leaf$

\ENDWHILE
\STATE Apply Gaussian blur: $A' \gets \text{GaussianBlur}(A_{DL,HR})$  
\STATE Downsample: $A_{DL,LR} \gets \text{BicubicDownsample}(A')$  

\end{algorithmic}
\end{algorithm}

The main idea of the dead leaves model is to generate an image by sequentially superimposing random shapes (called leaves) at random positions, thereby mimicking the process of dead leaves falling from a tree \cite{matheron1968modele, bordenave2006dead}. This model was in particular shown to be a powerful way to reproduce important statistics of natural images \cite{alvarez1999size, lee2001occlusion, gousseau2007modeling}. The shapes can be defined using random shape models, and their positions are given by a stationary Poisson point process (i.e., points uniformly distributed over the plane). The process is iterated until a stationary state is reached, which, in practice, can be achieved using perfect simulation techniques~\cite{kendall1999perfect}: each new shape is placed beneath the previous ones, until every pixel is covered by at least one leaf.

In our approach, we generate synthetic abundances using the dead leaves model, following the principle outlined above. More precisely, an abundance map is generated by sequentially superimposing shapes bearing random abundance values (in $\mathbb{R}^M$). When a given pixel is covered by a random shape, it is given the abundance values of this shape, therefore prescribing its $M$ material values in the abundance map (see Fig. \ref{fig_DL_exemple}). Given the urban context of HSIs in remote sensing, the use of rectangular leaves appears well-suited. Each leaf is characterized by a vector $(a, b, \theta, \textbf{v}, x, y)$, where $a$ and $b$ define the dimensions of the rectangle, $\theta$ its orientation, $\textbf{v} \in [0, 1]^M$ the abundance values assigned to the leaf and $(x, y)$ the spatial coordinates of its center. The dead leaves process is iterated until all pixels have been assigned at least one leaf.
In practice, the synthetic abundances $A_{DL,HR}$ are first generated at high resolution with spatial dimensions $(H,W,M)$. In this process, the values $a$ and $b$ are independently drawn from a uniform distribution over $[2 \times r, \min(H,W)/3]$, where $r = H/h = W/w$ denotes the super-resolution upscaling factor, ensuring variability in the leaf sizes. The value $\theta$ is uniformly drawn in the interval $[0^\circ, 45^\circ]$, capturing all possible orientations due to the symmetry of the shapes. The corresponding low-resolution abundances $A_{DL,LR}$ of size $(h,w,M)$ are subsequently obtained from $A_{DL,HR}$ through convolution with the PSF.

To preserve realistic abundance statistics within the dead leaves framework, the values $\mathbf{v}$ for each leaf in the synthetic abundances are sampled by randomly selecting pixels from $A_{LR}$ containing the material distribution $\mathbf{v}_{extract}$ (see Fig. \ref{fig_DL_exemple}). This strategy ensures realistic material proportions in the generated maps, as illustrated in Fig. \ref{fig_compare_DL}. An example comparing the power spectrum and per-endmember abundance histograms between real urban HSI abundances and dead-leaves synthetic abundances is shown in Fig. \ref{fig_PowSpec_Hist}. The detailed implementation for generating a single LR-HR synthetic abundance pair is provided in Algorithm \ref{alg:dead_leaves}. We subsequently use this algorithm to generate a large number of such pairs, thereby constituting our training dataset.

\subsection{Super-resolution Network}

We investigated the effectiveness of our synthetic training strategy with a state-of-the-art SISR super-resolution networks: MCNet (short for Mixed 3D/2D Convolutional Network). It is a supervised CNN initially developed for direct super-resolution of HSIs \cite{li2020mixed}. As the name suggests, it combines both 2D and 3D convolutional layers: the 2D layers focus on spatial feature extraction, while the 3D layers jointly capture spatial and spectral information. In our framework, the architectures is adapted to operate on abundance maps rather than hyperspectral images. Following the strategy of FFDNet \cite{zhang2018ffdnet}, we append a noise level map as an additional input channel, filled with the standard deviation of the actual Gaussian noise added to each sample. As a result, each network receives an input tensor with $M+1$ channels, where $M$ denotes the number of materials, which allows the model to explicitly account for varying noise levels and improves robustness during training. The performance of MCNet under this parameterization (referred to as MCNet-DL) will be evaluated through comparative experiments against state-of-the-art methods in Section~\ref{sub_sec_SOTA}. Moreover, since our framework is designed to be adaptable, we will also assess the impact of replacing the backbone network with other super-resolution architectures in the ablation studies presented in Sections~\ref{subsub_sec_Adaptability_DL}.

\subsection{Noise modeling}
\label{sub_sec_noise_model}
Adding noise to synthetic images is essential for enhancing the robustness of models trained on synthetic abundance. Assuming additive Gaussian white noise $N_{LR}$, and introducing the pseudo-inverse of the endmember matrix $S^{+}$, the observation model becomes:

\begin{multline}
HSI_{LR}(i,j,l) = \\
\sum_{m=1}^{M} \left[ A_{LR}(i,j,m) + \sum_{l'=1}^{L} N_{LR}(i,j,l') \cdot S^{+}(l',m) \right] \cdot S(m,l)
\label{eq:factorize_noise}
\end{multline}

This shows that the noise effect can be viewed as a direct perturbation $N_A = \sum_{l'=1}^{L} N_{LR}(i,j,l') \cdot S^{+}(l',m)$ applied to the abundance.
Rather than aiming to explicitly train the network for denoising, the objective here is to increase its robustness to slight input variations. To this end, we introduce a small additive Gaussian noise $N_{LR}$ characterized by a random standard deviation $\sigma$. To simulate varying noise intensities, the corresponding Peak Signal-to-Noise Ratio is drawn according to $PSNR(\sigma) \sim PSNR_{\text{max}} - \text{Exp}(\lambda)$, where $\sigma_{\text{max}}$ corresponds to the maximum noise level just above the quantization threshold of HSIs. In practice, we set $PSNR_{\text{max}} = 60\text{dB}$ and $\lambda = 5$. The resulting abundance-space noise $ N_A $ is then applied to half of the synthetic abundance $ A_{DL,LR} $. During training, the network learns to super-resolve both clean and slightly corrupted inputs (i.e., $ A_{DL,LR} $ or $ A_{DL,LR} + N_A $) into their corresponding high-resolution counterparts $ A_{DL,SR} $.

\section{Experiments}
\label{sec_Experiments}

This section presents a comprehensive evaluation of the proposed method through a series of experiments. First, in \ref{sub_sec_dataset}, we detail the experimental setup, including the datasets, preprocessing steps, training protocols, and evaluation metrics. Then, in \ref{sub_sec_SOTA}, we provide a comparative analysis with state-of-the-art supervised super-resolution methods, highlighting the effectiveness of our unsupervised approach. The design of our method is further justified in \ref{sub_sec_unmixing}, \ref{sub_sec_DL}, and \ref{sub_sec_noise}. Specifically, \ref{sub_sec_unmixing} explores the impact of different hyperspectral unmixing strategies through an ablation study, \ref{sub_sec_DL} analyzes the contributions and design choices of the dead leaves model for synthetic data generation, \ref{sub_sec_noise} investigates the effect of noise addition during training on network robustness, and \ref{sub_sec_cross_dataset} evaluates the generalization capability of our method through cross-dataset experiments using PRISMA hyperspectral data. Finally, \ref{sub_sec_PSF} studies how mismatches between the assumed and true PSF affect the super-resolved reconstructions, and \ref{sub_sec_efficiency} examines the computational efficiency of the proposed framework.

\subsection{Datasets \& Setup}
\label{sub_sec_dataset}

The proposed approach was evaluated on three widely used hyperspectral datasets:

\begin{itemize} 
    \item \textbf{Urban}: This dataset was captured by the Hyperspectral Digital Image Collection Experiment (HYDICE) sensor. It consists of $307 \times 307$ pixels across 210 spectral bands ranging from 400 to 2500 nm. After discarding noisy and corrupted bands, 162 bands are retained for evaluation.
    
    \item \textbf{Pavia University}: Acquired by the ROSIS-3 airborne optical sensor in 2003, this image comprises $610 \times 340$ pixels with a ground sampling distance (GSD) of 1.3 meters. It covers a spectral range of 430–840 nm distributed over 115 bands. Twelve bands affected by noise and water vapor absorption are removed. In this study, we focus on a $340 \times 340$ pixel region from the lower part of the image, using the remaining 103 bands.
    
    \item \textbf{Chikusei}: Captured by the VNIR-C sensor, the original image has a resolution of $2517 \times 2335$ pixels over 128 spectral bands. For our experiments, we randomly select a $500 \times 500$ pixel subregion.

\end{itemize}

To evaluate the performance of the proposed method, we apply down-sampling factors of $\times 2$, $\times 3$, and $\times 4$. The low-resolution hyperspectral images $HSI_{LR}$ are generated using a Gaussian blur followed by bicubic down-sampling. The standard deviation of the Gaussian filter is set to $\sigma = 2$, $3$, and $4$ for each respective down-sampling factor and the filter size is truncated to $6\sigma$ to comply with the Shannon–Nyquist sampling criterion.

For each dataset and each training phase, we generate 5,000 synthetic high-resolution abundance maps $A_{DL,HR}$ of size $W \times H \times 6$, where $W$ and $H$ denote the spatial dimensions of the corresponding dataset. The associated low-resolution abundances  $A_{DL,LR}$ are obtained by applying the appropriate PSF. Among these, 2,500 abundances are corrupted with noise following the protocol described in the previous section: for each sample synthetic image, a different noise level is randomly selected above the limit of quantification, following a decreasing exponential distribution.

The network is trained for 200 epochs on the corresponding synthetic dataset using the L1 loss function and the Adam optimizer, with a fixed learning rate of 0.0001. The other network-related parameters remain unchanged from the initial version proposed in MCNet. Performance is assessed using three metrics,  where $X$ denotes the reference high-resolution hyperspectral image and $Y$ denotes the reconstructed image:
\begin{itemize}

\item Peak Signal-to-Noise Ratio (PSNR), defined as  
  \begin{equation}
  \text{PSNR} = 10 \log_{10} \left( \frac{\text{MAX}^2}{\text{MSE}} \right)
  \end{equation}
  where $ \text{MAX} $ is the maximum possible pixel value and 
  \[
  \text{MSE} = \frac{1}{HWL} \sum_{i=1}^{H} \sum_{j=1}^{W} \sum_{l=1}^{L} \left( X(i,j,l) - Y(i,j,l) \right)^2
  \]
  with $H$, $W$, and $L$ denoting the height, width, and number of spectral bands, respectively.

\item Spectral Angle Mapper (SAM), which computes the average spectral angle between corresponding spectral vectors in $X$ and $Y$:
  \begin{equation}
  \text{SAM} = \frac{1}{HW} \sum_{i=1}^{H} \sum_{j=1}^{W} \cos^{-1} \left( \frac{ \sum_{l=1}^{L} X(i,j,l) \cdot Y(i,j,l) }{ \|X(i,j,.)\|_2 \cdot \|Y(i,j,.)\|_2 } \right)
  \end{equation}
  where $X(i,j,.)$ and $Y(i,j,.)$ denote the spectral vectors at pixel location $(i,j)$.

\item Error Relative Global Adimensional Synthesis (ERGAS), given by  
  \begin{equation}
  \text{ERGAS} = 100 \cdot \frac{d}{r} \cdot \sqrt{ \frac{1}{l} \sum_{l=1}^{L} \left( \frac{\text{RMSE}_l}{\mu_l} \right)^2 }
  \end{equation}
where 
  \[
  \text{RMSE}_l = \sqrt{ \frac{1}{HW} \sum_{i=1}^{H} \sum_{j=1}^{W} \left( X(i,j,l) - Y(i,j,l) \right)^2 }
  \]
and $\mu_l$ is the mean value of band $l$ in $Y$, while $\frac{d}{r}$ is the spatial resolution ratio between LR and HR images.
\end{itemize}

Note that the ablation studies in \ref{sub_sec_unmixing}, \ref{sub_sec_DL} and \ref{sub_sec_noise} are only carried out on the Urban dataset, focusing on the most challenging super-resolution setting, namely $\times 4$.

\subsection{Comparison with the State-of-the-Art Methods}
\label{sub_sec_SOTA}

\begin{figure*}
    \centering
    \includegraphics[width=0.85\linewidth]{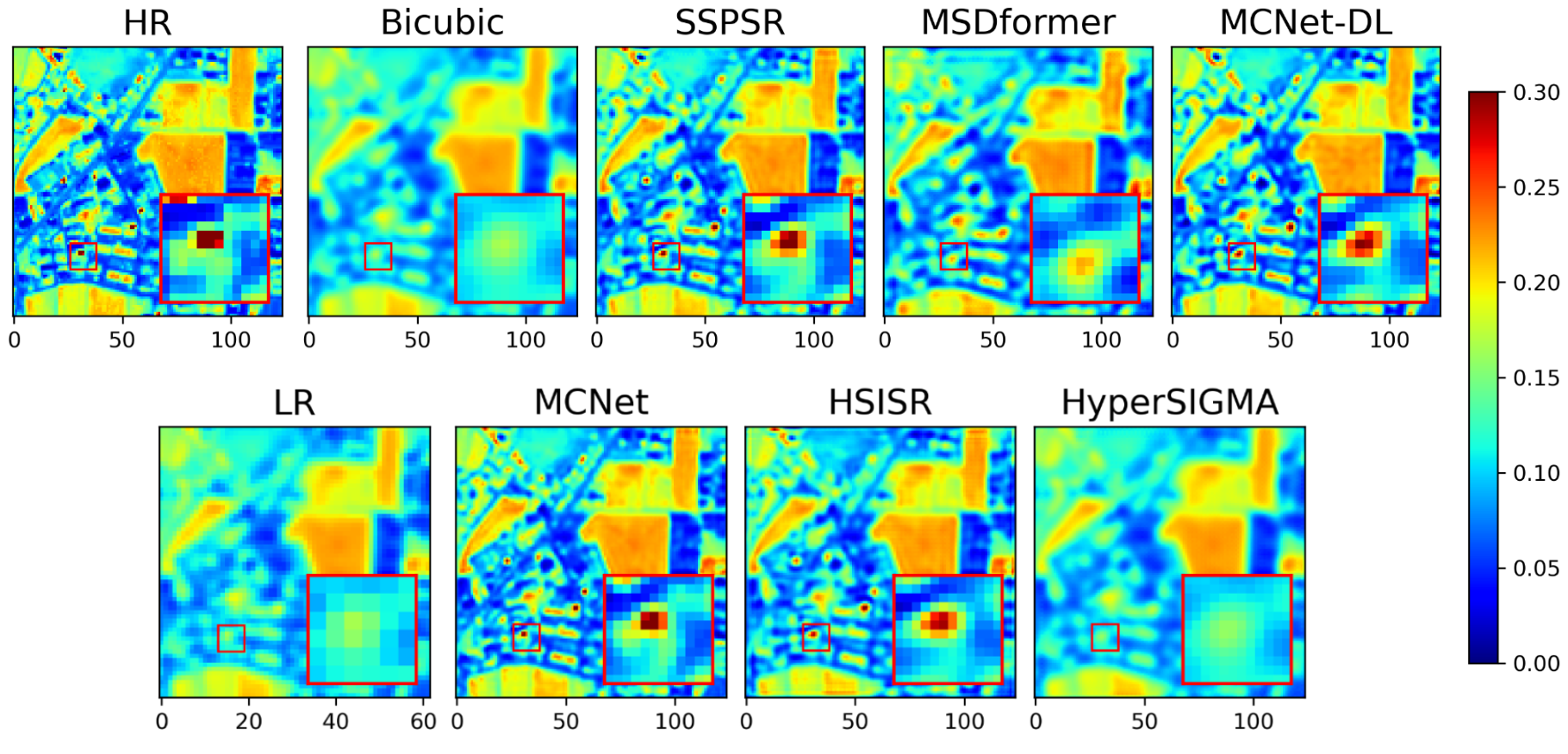}
    \caption{Visual comparison at band 100 of a \textbf{Chikusei} patch with a $\times 2$ scaling factor: high-resolution reference (HR), low-resolution (LR), and super-resolved (SR) images generated using bicubic interpolation, MCNet, SSPSR, HSISR, MSDformer, HyperSIGMA, MSDformer-DL and MCNet-DL.}
    \label{fig_Chikusei_X2}
\end{figure*}

\begin{figure*}
    \centering
    \includegraphics[width=0.85\linewidth]{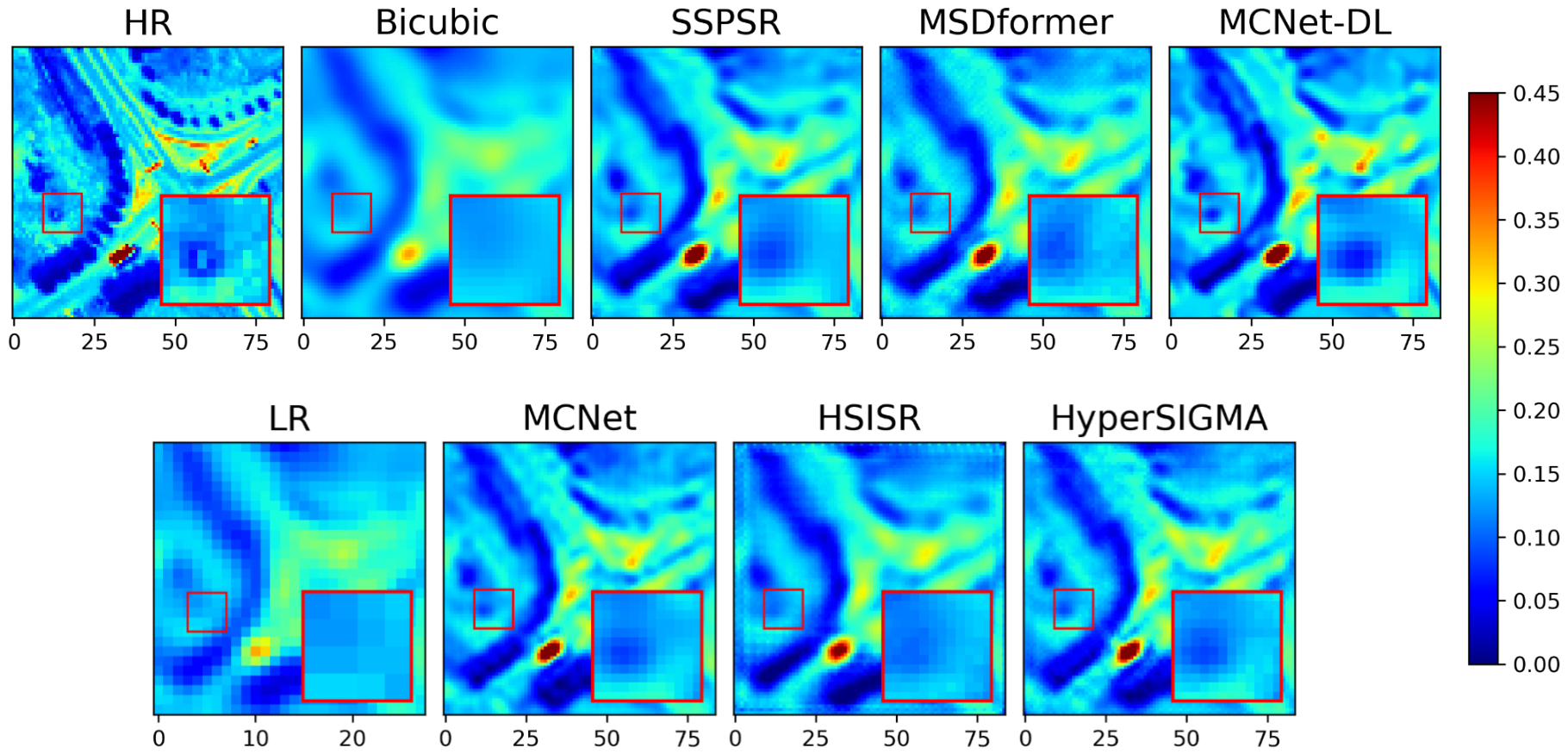}
    \caption{Visual comparison at band 50 of a \textbf{Pavia University} patch with a $\times 3$ scaling factor: high-resolution reference (HR), low-resolution (LR), and super-resolved (SR) images generated using bicubic interpolation, MCNet, SSPSR, HSISR, MSDformer, HyperSIGMA, MSDformer-DL and MCNet-DL.} 
    \label{fig_PaviaU_X3}
\end{figure*}

\begin{figure*}
    \centering
    \includegraphics[width=0.85\linewidth]{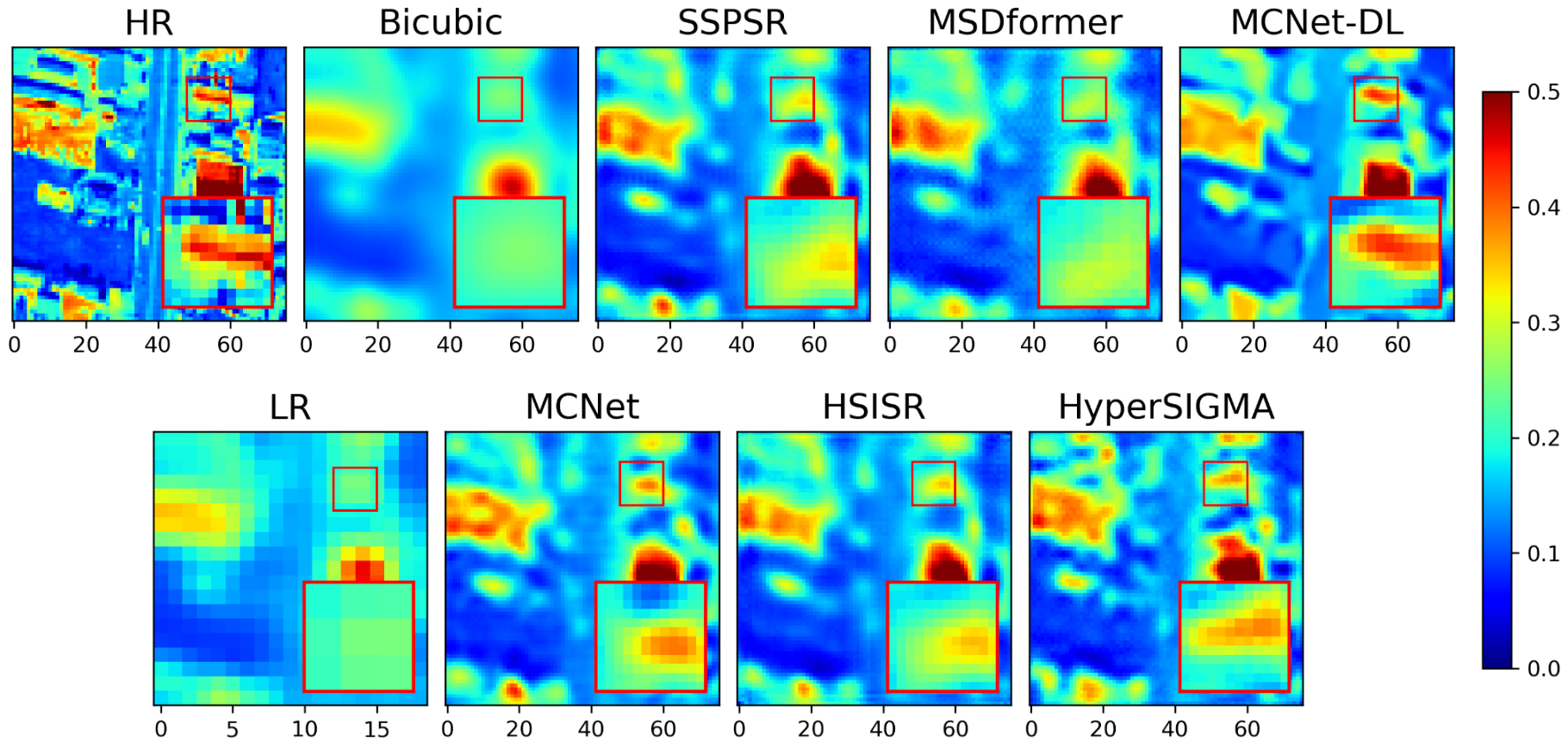}
    \caption{Visual comparison at band 50 of a \textbf{Urban} patch with a $\times 4$ scaling factor: high-resolution reference (HR), low-resolution (LR), and super-resolved (SR) images generated using bicubic interpolation, MCNet, SSPSR, HSISR, MSDformer, HyperSIGMA, MSDformer-DL and MCNet-DL. }
    \label{fig_Urban_X4}
\end{figure*}

\begin{table*}
\centering
\caption{Average super-resolution performance across scales ($\times2$, $\times3$, $\times4$) on three hyperspectral datasets, comparing SOTA supervised methods with our unsupervised MCNet-DL trained solely on synthetic data.}
\resizebox{\linewidth}{!}{
\begin{tabular}{ll|ccc|ccc|ccc}

 &  & \multicolumn{3}{c|}{\textbf{Urban}} & \multicolumn{3}{c|}{\textbf{Pavia University}} & \multicolumn{3}{c}{\textbf{Chikusei}} \\

\textbf{Scale} & \textbf{Method} & mPSNR $\uparrow$ & mSAM $\downarrow$ & mERGAS $\downarrow$ & mPSNR $\uparrow$ & mSAM $\downarrow$ & mERGAS $\downarrow$ & mPSNR $\uparrow$ & mSAM $\downarrow$ & mERGAS $\downarrow$ \\
\hline 
\multirow{7}{*}{$\times2$}
& Bicubic        & 26.46 & 14.79 & 7.87 & 29.10 & 11.08 & 5.91 & 34.39 & 13.24 & 8.08\\
& MCNet         & \underline{30.05}&\underline{9.85}&\underline{5.27}& \underline{33.02} & \textbf{7.11} & \textbf{3.93} &\textbf{39.63} & \textbf{7.36} & \textbf{4.60} \\
& SSPSR         & 29.43 & 10.55 & 5.62 & 31.59 & 8.34 & 4.61 & 38.79 & 8.22 & 5.21 \\
& HSISR         & 29.27 & 10.29 & 5.50 & 27.98 & 11.66 & 7.10 & 32.78 & 13.99 & 10.66 \\
& HyperSIGMA         & 29.73& 10.45& 5.44& 30.20 & 8.59&4.09&33.44&16.00&9.37\\
& MSDformer      & 29.34&10.98&5.71& 26.29&15.62&7.76 & 33.65&15.63&9.15 \\
& MCNet-DL      & \textbf{30.47}&\textbf{9.33}&\textbf{4.93}& \textbf{33.09} & \underline{7.30} & \underline{4.02} & \underline{39.29} & \underline{7.86} & \underline{4.78} \\
\hline
\multirow{7}{*}{$\times3$}
& Bicubic       & 24.99 & 17.54 & 9.30 & 27.24 & 13.60 & 7.15 & 32.51 & 16.41 & 9.93\\
& MCNet         & \underline{27.12} &\underline{13.67} &\underline{7.28} & \underline{29.76}& \underline{10.23} & \underline{5.49} & 35.25 & 12.14 & 7.41 \\
& SSPSR         & 26.51 & 14.70 & 7.52 & 29.37 & 10.78 & 5.88 & \underline{35.36} & \underline{11.78} & \underline{7.31} \\
& HSISR         & 26.79 & 14.41 & 7.75 & 28.14 & 12.15 & 6.76 & 32.29 & 16.09 & 10.71 \\
& HyperSIGMA         & 26.67& 14.84&7.67 & 29.10 & 11.66& 7.87 & 32.69&17.43&10.06\\
& MSDformer      & 26.48&15.17&7.82 & 28.88 & 11.97 &6.03 &32.39&17.83&10.26 \\
& MCNet-DL      & \textbf{28.05} & \textbf{12.19} & \textbf{6.41} & \textbf{30.75} & \textbf{9.42} & \textbf{5.03} & \textbf{36.53} & \textbf{10.67} & \textbf{6.52} \\
\hline
\multirow{7}{*}{$\times4$}
& Bicubic       & 24.31 & 19.09 & 10.14 & 26.33 & 15.21 & 7.93 & 31.52 & 18.36 & 11.02 \\
& MCNet         & 25.76 & 16.26 & 8.61 & \underline{28.19} & \underline{12.17} & \underline{6.46} & 33.28  & 15.16 & 9.16 \\
& SSPSR         & 25.13 & 17.08 & 8.84 & 27.64 & 12.53 & 6.79 & \underline{33.40} & \underline{15.01} & \underline{8.95} \\
& HSISR         & 25.41 & 17.01 & 8.84 & 25.74 & 15.12 & 9.05 & 29.42 & 21.01 & 14.86 \\
& HyperSIGMA         & \underline{26.02} &  \underline{16.14} & \underline{8.31} &26.07 & 15.93& 7.96&32.34&17.85&10.26\\
& MSDformer      & 25.03&18.14&9.32 & 26.17&16.11&8.04 & 31.47&19.76&11.23\\
& MCNet-DL      & \textbf{26.54} & \textbf{14.71} & \textbf{7.67} & \textbf{29.34} & \textbf{11.06} & \textbf{5.84} & \textbf{34.99} & \textbf{12.92} & \textbf{7.85} \\

\end{tabular}
}
\label{tab_SOTAs_results}
\end{table*}

 We compare our method, MCNet-DL, trained exclusively on a fully synthetic dataset, against state-of-the-art (SOTA) SISR methods: MCNet \cite{li2020mixed}, SSPSR \cite{jiang2020learning}, HSISR \cite{li2022hyperspectral}, MSDformer \cite{chen2023msdformer} and a method based on a very recent foundation model, HyperSIGMA \cite{wang2025hypersigma}. Bicubic interpolation is also included as a baseline. For the SOTA methods, we adopt the default hyperparameters recommended by the original authors, without additional tuning. It is worth noting that all these SOTA methods are supervised, which makes the comparison more favorable to them, since they benefit from access to paired training data, unlike our approach.

To construct the training data, the dataset considered is divided into 16 non-overlapping patches. For each SOTA method, 16 independent training sessions are conducted: in each session, one patch is used exclusively for evaluation, while the remaining 15 serve as training data. This procedure, which relies on an artificial partitioning of the data, offers favorable learning conditions for supervised methods and thus favors them. However, such a setup is unrealistic in real-world applications.

In contrast, our method is trained solely on the synthetic dead leaves dataset composed of 5,000 abundances. Although our unsupervised approach allows evaluation over the entire image, we adopt the same patch-based evaluation protocol for fair comparison with SOTA methods. It is important to note that this choice modifies the quantitative results compared to those reported in the ablation studies.

Table \ref{tab_SOTAs_results} summarizes the average performance over the 16 evaluation patches for each method, each metric, and each studied scale factor. Similarly, Figures \ref{fig_Chikusei_X2}, \ref{fig_PaviaU_X3}, and \ref{fig_Urban_X4} provide visual comparisons on a representative patch for each scale factor. Quantitatively, MCNet-DL achieves state-of-the-art results in the majority of cases across all datasets and scale factors. While MCNet-DL slightly underperforms MCNet on the $\times2$ scale for Chikusei and Pavia University, particularly on mSAM and mERGAS, it consistently outperforms all other methods. At higher magnification factors ($\times3$ and $\times4$, which actually corresponds to interesting zoom factors for remote sensing), MCNet-DL clearly yields the best results across all metrics and datasets. It is important to highlight that MCNet, SSPSR, HSISR, MSDformer, and HyperSIGMA benefit from a highly favorable and supervised training setup, relying on real data from the same scene being evaluated, which constitutes an unrealistic condition in practical applications. In contrast, MCNet-DL is trained in a fully unsupervised manner. Among the results for $\times3$ and $\times4$ scales, it is worth noting in particular that, MCNet-DL clearly surpasses MCNet across all metrics and datasets, highlighting the efficiency of our synthetic-data-based approach. Visually, MCNet-DL reconstruct more accurate spatial details and preserve spectral information better than all other methods, with improvements especially noticeable in high-frequency regions and complex scenes, as illustrated in the qualitative examples.

\subsection{Hyperspectral Unmixing}
\label{sub_sec_unmixing}

\begin{table}
\centering
\caption{Super-resolution performance ($\times 4$) using different methods of unmixing/decomposition on the Urban dataset with $M=6$.}
\begin{tabular}{lccc}
\toprule
\textbf{Decomposition method} & \textbf{PSNR} $\uparrow$ & \textbf{SAM} $\downarrow$ & \textbf{ERGAS} $\downarrow$ \\
\midrule
endmembers GT       & \textbf{26.19} & \textbf{14.89} & \textbf{7.44} \\
MinVol                  & \underline{25.75} & \underline{15.73} & \underline{7.85} \\
CNNAEU                  & 24.83 & 18.23 & 9.19 \\
PCA         & 14.55 & 67.33 & 25.70 \\
\bottomrule
\end{tabular}
\label{tab_unmixing_results}
\end{table}

\begin{figure}
    \centering
    \includegraphics[width=1\linewidth]{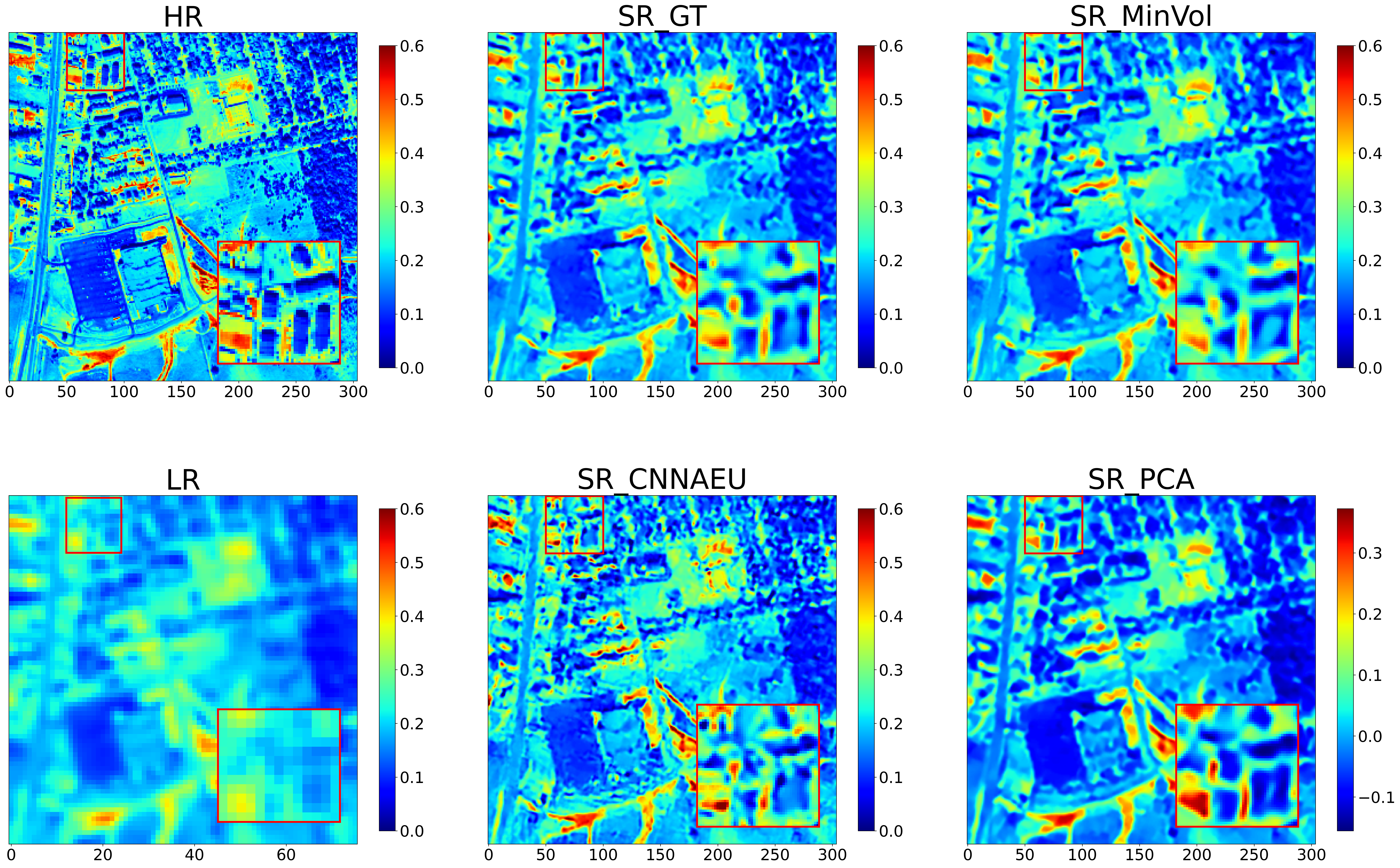}
    \caption{Visual comparison at band 100: high-resolution reference (HR), Low-resolution (LR), and super-resolved (SR) images obtained using endmembers from ground truth (GT), MinVol, CNNAEU, and PCA-based decomposition.}
    \label{fig_unmixing_results}
\end{figure}

This section aims to highlight the importance of hyperspectral unmixing in the proposed framework, using the Urban dataset as a benchmark. To this end, we compare several endmember extraction strategies, including a simple Principal Component Analysis (PCA) decomposition, a classical blind unmixing method known as MinVol \cite{gillis2014successive}, and a deep learning-based approach, CNNAEU, proposed by Pálsson et al. \cite{palsson2020convolutional}. We also include the commonly used ground-truth endmembers, annotated by Zhu \cite{zhu2017hyperspectral}, as a reference.

The Urban dataset is particularly suited for this analysis, as it includes ground-truth versions with different numbers of endmembers: $M=4$, $M=5$, and $M=6$. In this study, we focus on the case with $M=6$ endmembers, as it better preserves spectral diversity and reduces compression artifacts typically associated with lower-dimensional representations.

As outlined above, regardless of the unmixing method employed, we retain only the extracted spectral signatures and apply a least-squares procedure to estimate the corresponding abundance. This approach limits reconstruction error and facilitates the analysis of spatially correlated noise in the abundance domain, enabling a fair and consistent evaluation of super-resolution performance across methods.

Table~\ref{tab_unmixing_results} summarizes the quantitative super-resolution performance achieved with different endmember extraction strategies, while Figure~\ref{fig_unmixing_results} provides a visual illustration of the corresponding reconstruction results. As expected, using ground-truth endmembers gives the best results for all measurements. Among blind unmixing methods, the MinVol approach achieves performance close to the ground truth, demonstrating its relevance. Although CNNAEU still outperforms PCA, its accuracy remains slightly lower than that of GT and MinVol, and the reconstruction exhibits more noticeable artefacts. In contrast, the PCA-based decomposition yields significantly poorer results, with dynamics that are markedly shifted and tend to produce negative values. Moreover, its components often lack clear physical or visual relevance, making them less informative for interpretation. These results underscore the value of dedicated unmixing techniques in extracting meaningful spectral signatures. Since ground-truth endmembers are not always accessible, we rely on the MinVol method for spectral extraction.

\subsection{Ablation studies on dead leaves}
\label{sub_sec_DL}

\begin{table}
\centering
\caption{Super-resolution performance ($\times 4$) using different synthetic abundance generation strategies on the Urban dataset with $M=6$.}
\begin{tabular}{lccc}
\toprule
\textbf{SR strategy} & \textbf{PSNR} $\uparrow$ & \textbf{SAM} $\downarrow$ & \textbf{ERGAS} $\downarrow$ \\
\midrule
MCNet-DL        & \textbf{25.75} & \textbf{15.73} & \textbf{7.85} \\
MCNet-Nat       & 20.91 & 26.76 & 13.54 \\
MCNet-DirichletDL    & 24.47 & 18.51 & 9.26 \\
\bottomrule
\end{tabular}
\label{tab_DL_comparison}
\end{table}

In this section, we present ablation studies for the dead leaves model, focusing on three key aspects: (1) the relevance of using dead leaves, (2) the importance of preserving the material distribution from real low-resolution abundance $A_{LR}$, and (3) the model’s adaptability across different super-resolution network architectures. Table \ref{tab_DL_comparison} summarizes all the tests performed for the section ablation studies.

\subsubsection{The relevance of using dead leaves}

The idea is to demonstrate the relevance of the dead leaves model for generating synthetic abundance. To assess this, by using MCNet, we compare DL abundances (denoted as MCNet-DL) to an alternative approach that uses natural image statistics. Specifically, we leverage the DIV2K dataset \cite{agustsson2017ntire}, which contains a wide range of high-resolution natural images. By extracting random patches from these images and interpreting them as proxy abundance distributions (denoted as MCNet-Nat).

Quantitative results from Table \ref{tab_DL_comparison} clearly show that MCNet-DL significantly outperforms MCNet-Nat across all evaluation metrics. The dead leaves model not only provides higher spatial fidelity (as reflected by the PSNR) but also better preserves spectral characteristics, as indicated by the lower SAM and ERGAS values. These results support the relevance of the DL model, whose structured and layered composition produces training data that more faithfully captures the physical and spatial constraints inherent in real hyperspectral mixtures. This leads to more accurate and robust super-resolution performance. In comparison, although natural image statistics offer high variability, they do not reflect the underlying semantics of abundance mixing, ultimately resulting in less effective learning.

\subsubsection{The material distribution}

We now assess the importance of preserving the material distribution observed in real data. To this end, we compare MCNet-DL, which samples each leaf’s value $V$ from real low-resolution abundance $A_{LR}$, to MCNet-DirichletDL, a variant where $V$ is instead drawn from a uniform Dirichlet distribution $\mathrm{Dir}(1,\dots,1)$. In this setting, for each new leaf, an $N$-dimensional abundance vector $\mathbf{\textbf{v}} \sim \mathrm{Dir}(1,\dots,1)$ is sampled, and its $m$\textsuperscript{th} component is assigned to the leaf at layer $m$. All geometric parameters $(a,b,\theta,x,y)$ remain identical between the two methods.

Results in Table \ref{tab_DL_comparison} show that MCNet-DL consistently outperforms MCNet-DirichletDL across all metrics. These results clearly demonstrate that leveraging the true material distribution from $A_{LR}$ leads to more realistic and effective training data, whereas sampling from a uniform Dirichlet prior introduces unlikely combinations of materials that hinder super-resolution performance.

\subsubsection{Adaptability to various super-resolution networks}
\label{subsub_sec_Adaptability_DL}

\begin{table}
\centering
\caption{Super-resolution performance ($\times 4$) using different super-resolution networks on the Urban dataset with $M=6$.}
\begin{tabular}{lccc}
\toprule
\textbf{SR strategy} & \textbf{PSNR} $\uparrow$ & \textbf{SAM} $\downarrow$ & \textbf{ERGAS} $\downarrow$ \\
\midrule
MCNet-DL        & \textbf{25.75} & \textbf{15.73} & \textbf{7.85} \\
MSDformer-DL   & 25.26 & 16.32 & 8.45 \\
RDN-DL   & 23.83 & 19.91 & 9.85 \\
\bottomrule
\end{tabular}
\label{tab_network_comparison}
\end{table}

One significant advantage of the dead leaves dataset is its versatility across various super-resolution architectures. In our experiments, the synthetic training strategy has been instantiated with MCNet-DL. To further assess this adaptability, we replace MCNet with MSDformer and RDN. MSDformer \cite{chen2023msdformer} adopts a multiscale deformable Transformer architecture that leverages spatial–spectral self-attention and deformable sampling to capture long-range dependencies and handle complex hyperspectral structures. Similarly to MCNet, the input layer of MSDformer can be easily modified to accommodate $M + 1$ input channels corresponding to the abundance maps used in our framework. In contrast, RDN \cite{zhang2018residual} is a general-purpose super-resolution model originally developed for natural images. It employs a residual dense architecture and can naturally process single-channel grayscale images, which makes it straightforward to adapt for $M + 1$ input channels as required by our method. The combination of MSDformer and the proposed framework is referred to as MSDformer-DL, while the configuration using RDN is denoted as RDN-DL.

According to the results reported in Table~\ref{tab_network_comparison}, MCNet-DL achieves the best overall performance across all metrics, confirming the effectiveness of our synthetic training framework with hyperspectral-specific architectures. MSDformer-DL follows closely, showing only a slight performance drop compared to MCNet-DL. The consistent results of MSDformer-DL demonstrate that transformer-based models can also benefit from the dead leaves training strategy, highlighting its robustness across different architectural paradigms. RDN-DL, while performing slightly below the hyperspectral backbones, still delivers competitive results despite being originally designed for natural-image super-resolution. This gap is expected, since RDN, while effectively capturing spatial structures, does not account for the distribution and interactions of materials across the abundance maps.

Overall, these results demonstrate that the dead leaves model offers a flexible and reliable basis for hyperspectral super-resolution tasks, regardless of the particular architecture used.

\paragraph*{In summary} The dead leaves model proves to be a powerful and versatile tool for synthetic data generation in hyperspectral image super-resolution. It not only generates physically realistic spatial structures that better match real abundance but also preserves meaningful material statistics and supports diverse network architectures. These advantages collectively enable improved model performance, robustness, and generalization.

\subsection{Discussion on Noise and network training robustness}
\label{sub_sec_noise}

\begin{table}
\centering
\caption{Super-Resolution Performance ($\times 4$) under Different Training Configurations on Clean and Noisy Test Data (Urban dataset, $M=6$)}
\begin{tabular}{llccc}
\toprule
\textbf{Test Dataset} & \textbf{DL Configuration} & \textbf{PSNR} $\uparrow$ & \textbf{SAM} $\downarrow$ & \textbf{ERGAS} $\downarrow$ \\
\midrule
\multirow{4}{*}{Urban}
& Clean & \textbf{26.10} & \textbf{15.13} & \textbf{7.58} \\
& Noisy & 25.45 & 16.44 & 8.19 \\
& HalfMix & \underline{25.85} & \underline{15.62} & \underline{7.84}\\
& StdAware & 25.75 & 15.73 & 7.85\\
\midrule
\multirow{4}{*}{Urban + 60dB}
& Clean & 16.94 & 39.42 & 22.45\\
& Noisy & \underline{25.43} & \underline{16.49} & \underline{8.22}\\
& HalfMix & 23.86 & 20.02 & 9.82\\
& StdAware & \textbf{25.57} & \textbf{16.12} & \textbf{8.04}\\

\midrule
\multirow{4}{*}{Urban + 30dB}
& Clean &  0.96 & 78.15 & 142.54\\
& Noisy & \textbf{24.19} & \textbf{18.56} & \textbf{9.58}\\
& HalfMix &16.92 & 39.47 & 20.91 \\
& StdAware & \underline{24.03} & \underline{18.91} & \underline{9.76}\\

\bottomrule
\end{tabular}
\label{tab_Noise}
\end{table}

\begin{figure*}
    \centering
    \includegraphics[width=0.85\linewidth]{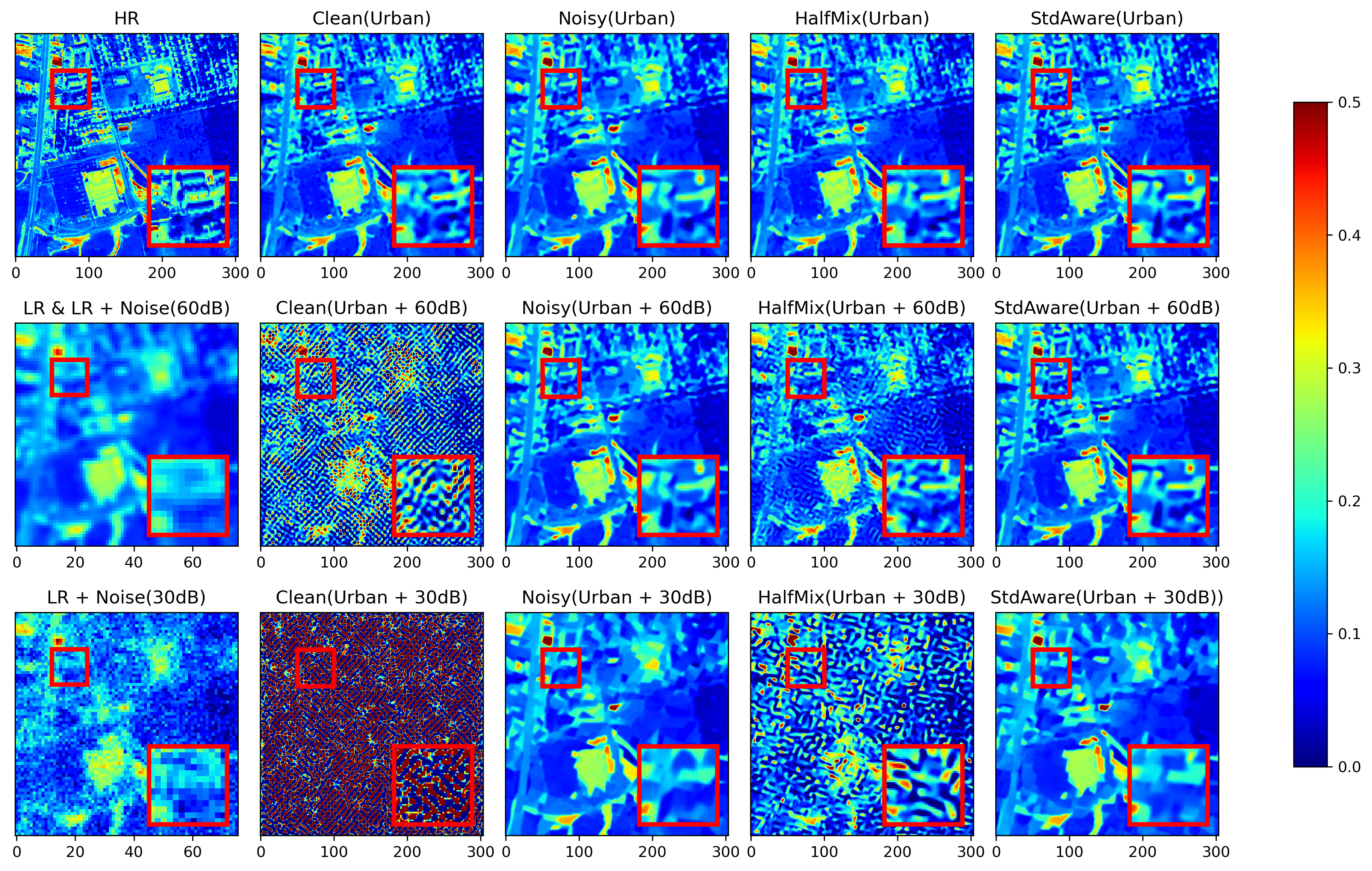}
    \caption{Visual Comparison at band 50 of Super-Resolved results under Different Training Configurations (Urban dataset, $M=6$, $\times 4$)}
    \label{fig_Noise}
\end{figure*}

We address the question of noise by showing that adding a small amount of noise during training is essential for network stability. Without it, models trained exclusively on clean synthetic data become overly sensitive to minor fluctuations at test time, especially when applied to slightly noisy real-world abundance. This leads to the appearance of ringing artifacts, as the network mistakenly treats these small variations as meaningful details to super-resolve. By introducing noises during training, we regularize the network’s behavior and effectively eliminate such artifacts, resulting in more robust and reliable reconstructions.

To investigate this effect, 3 test datasets were prepared. The first is the Urban dataset, which has been used throughout the ablation studies, with a scale factor of $\times 4$ and $M=6$ endmembers. The second is a noisy version of the Urban dataset, where additive Gaussian noise corresponding to a PSNR of $60\mathrm{dB}$ was applied to the low-resolution HSI. This noise level is extremely low, close to the quantization limit, and remains imperceptible to the human eye. To further assess the robustness of the super-resolution model under higher noise conditions, a third version of the Urban dataset was generated by introducing an additive Gaussian noise yielding a PSNR of $30\,\mathrm{dB}$ in the low-resolution HSI. This level corresponds to one of the highest noise intensities according to the noise distribution presented in Section~\ref{sub_sec_noise_model}, and is therefore among the least represented cases in the training distribution.

Regarding the training dataset configuration, 4 networks were trained independently and under identical conditions on different datasets. The first dataset consists exclusively of noiseless DL abundances (Clean). The second dataset contains the same DL abundances, but all are corrupted with random Gaussian noise generated as described in Section \ref{sub_sec_noise_model}, with an effective PSNR level above the quantization threshold at the hyperspectral image level (Noisy).  
 The third dataset is a balanced mixture composed of 50\% Clean and 50\% Noisy samples (HalfMix). Finally, the fourth dataset includes the same abundances as HalfMix, with an additional noise level map concatenated to the abundance, increasing the number of channels to $M+1$, as described in previous sections (StdAware).

An analysis of both the visual results (Fig.\ref{fig_Noise}) and the quantitative metrics (Tab.\ref{tab_Noise}) reveals that, for the Urban dataset without noise addition, training exclusively on Clean data yields the best results, while the Noisy training exhibits a slight performance decline, consistent with the nature of the respective datasets. However, when evaluated on the \textit{Urban + 60dB} test set, the detrimental effects of noise become evident: networks trained on Clean data display pronounced artifacts, and the HalfMix configuration only partially mitigates these effects, leading to a noticeable degradation in performance. These observations underscore the importance of introducing noise during training to enhance robustness. This trend is confirmed for the \textit{Urban + 30 dB} dataset, which represents one of the highest noise levels generated by the data synthesis process: the impact of noise becomes dramatic. Networks trained on Clean and HalfMix data exhibit poor performance, with pronounced ringing artifacts and severe distortions. In contrast, the Noisy and StdAware models manage to maintain reasonable performance under such challenging conditions, highlighting the importance of incorporating variable noise levels during training to achieve robust super-resolution in heavily degraded scenarios.

As a conclusion of this ablation study about noise, training exclusively on clean data leads to a low robustness of the method, while training only on noisy data yield a significant loss of performance when applied to clean images. Consequently, the StdAware configuration emerges as the most effective compromise, offering improved resilience to noise while preserving high reconstruction quality on noiseless data.

\subsection{Discussion on PSF}
\label{sub_sec_PSF}

\begin{figure*}
    \centering
    \includegraphics[width=0.85\linewidth]{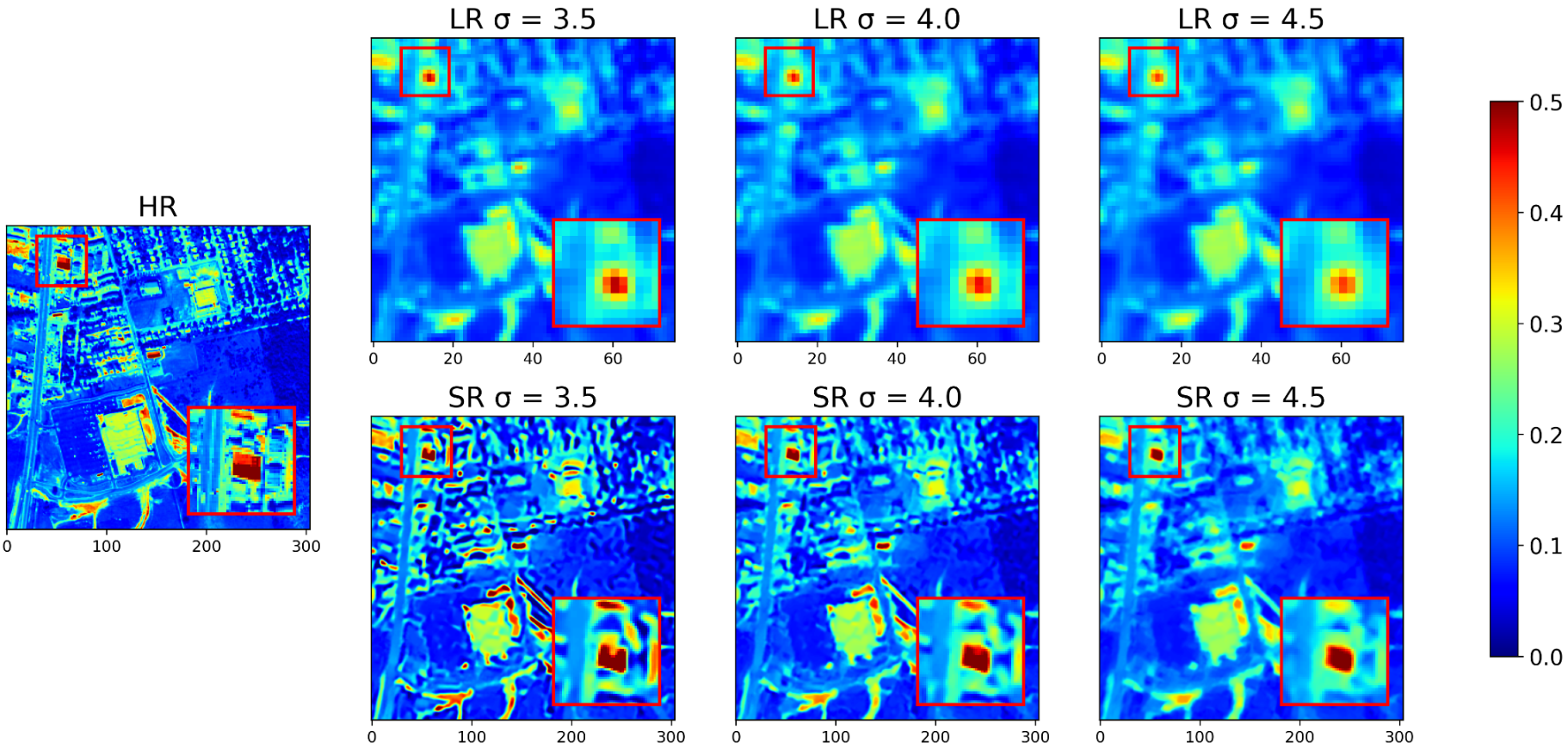}
    \caption{Visual Comparison at band 50 of PSF Mismatch on Super-Resolution Performance (Urban dataset, $M=6$, $\times 4$)}
    \label{fig_PSF_Fig}
\end{figure*}

\begin{figure*}
    \centering
    \includegraphics[width=0.85\linewidth]{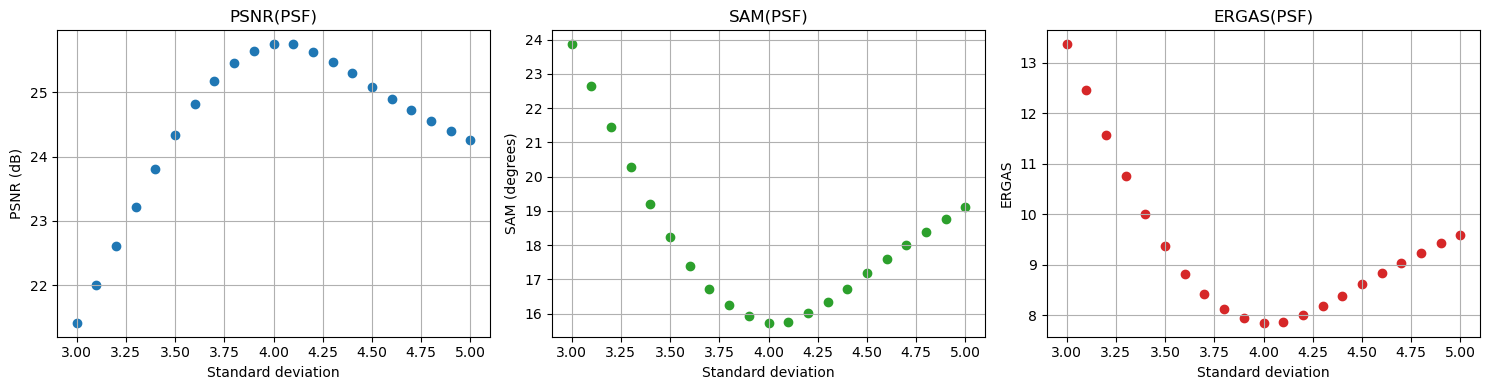}
    \caption{Overall Impact of PSF Mismatch on Super-Resolution Performance (Urban dataset, $M=6$, $\times 4$)}
    \label{fig_PSF_metrics}
\end{figure*}

This section investigates the sensitivity of the proposed method to inaccuracies in the PSF used for DL data generation. As done in most HSI SISR methods, our approach assumes that the PSF is known a priori, and, in the experimental setup of this paper, a controlled simulated scenario is considered in which the PSF is modeled as a Gaussian blur followed by bicubic downsampling. The objective here is to study the impact of PSF estimation errors in a general setting, without introducing sensor-specific characteristics tied to a particular hyperspectral instrument.

As a remainder, our network is trained using synthetic data generated with a Gaussian kernel of standard deviation $\sigma = 4$. To assess robustness to PSF mismatch, the trained MCNet-DL model is here evaluated on Urban hyperspectral images whose PSF is simulated with different Gaussian standard deviations ranging from $\sigma = 3$ to $\sigma = 5$, with a step of 0.1. The evolution of PSNR, SAM and ERGAS with respect to $\sigma$ is reported in Fig.~\ref{fig_PSF_metrics}, while qualitative results for $\sigma = 3.5$, $\sigma = 4.0$, and $\sigma = 4.5$ are shown in Fig.~\ref{fig_PSF_Fig}.

The quantitative curves exhibit a clear optimum around the true training value $\sigma = 4$, with PSNR peaking and both SAM and ERGAS reaching their best values in a narrow neighborhood of this setting. Moderate deviations of the PSF (e.g., $\sigma$ between 3.6 and 4.4) lead to only a limited degradation in performance, indicating a certain tolerance of the model to small PSF mismatches. However, larger discrepancies such as $\sigma = 3.0$ or $\sigma = 5.0$ result in a marked drop in PSNR and a significant increase in SAM and ERGAS, which is consistent with the visual inspection in Fig.~\ref{fig_PSF_Fig}, where blurring mismatches produce noticeable loss of details in the reconstructed images for $\sigma=4.5$ and strong artefacts for $\sigma = 3.5$.

\subsection{Cross-Dataset Comparison}
\label{sub_sec_cross_dataset}

To assess the generalization capability of our method across distinct images from the same hyperspectral sensor, we utilize hyperspectral data from the Precursore Iperspettrale della Missione Operativa (PRISMA) mission, launched by the Italian Space Agency in 2019 \cite{cogliati2021prisma}. Specifically, focusing on urban remote sensing scenes, we conduct our analysis using two PRISMA images: one of Athens and one of Paris, each comprising $230$ spectral bands. We apply the full pipeline from hyperspectral unmixing and synthetic abundance generation to network training on both images, denoted as MCNet-DL (Athens) and MCNet-DL (Paris). Both models are then evaluated on the Paris image with a $\times 4$ scaling factor, simulated using the same PSF as in previous experiments. Additionally, mirroring the protocol in Section \ref{sub_sec_SOTA}, we train supervised baselines MCNet, SSPSR, HSISR, HyperSIGMA, and MSDformer on non-overlapping patches only from PRISMA-Athens and test them on PRISMA-Paris.

The performance results on the PRISMA Paris image are reported in Table~\ref{tab_PRISMA_comparison}. Focusing only on the models trained with the Athens data, MCNet-DL (Athens) achieves the best performance among all compared methods, outperforming existing supervised approaches. This finding is consistent with the state-of-the-art comparison discussed in Section~\ref{sub_sec_SOTA}. Furthermore, the very close results obtained between MCNet-DL (Paris) and MCNet-DL (Athens) demonstrate the potential of our method to perform cross-dataset generalization when applied to data acquired by the same hyperspectral sensor.

\begin{table}
\centering
\caption{Cross-dataset super-resolution performance ($\times 4$) on PRISMA hyperspectral data, with all methods trained on PRISMA Athens and evaluated on PRISMA Paris, except MCNet-DL (Paris) trained and tested on PRISMA Paris.}
\begin{tabular}{lccc}
\toprule
\textbf{Method} & \textbf{PSNR} $\uparrow$ & \textbf{SAM} $\downarrow$ & \textbf{ERGAS} $\downarrow$ \\
\midrule
Bicubic & 27.95&12.43&7.22  \\
MCNet &30.79&11.51&6.62 \\
SSPSR  & 28.12&16.85&9.13  \\
HSISR  & 29.67&14.09&6.41  \\
HyperSIGMA  & 27.53&18.37&8.27 \\
MSDformer  & 29.86&14.23&7.52 \\
MCNet-DL (Athens)  &  \underline{32.33}&\underline{8.47}&\underline{6.21}  \\
MCNet-DL (Paris)  &\textbf{32.37}&\textbf{8.38}&\textbf{6.19} \\
\bottomrule
\end{tabular}
\label{tab_PRISMA_comparison}
\end{table}

\subsection{Computational Efficiency}
\label{sub_sec_efficiency}

\begin{table}
\centering
\caption{Computational cost for different hyperspectral SR methods on the Urban dataset with $M=6$ (scale $\times 4$).}
\begin{tabular}{lcccc}
\toprule
\textbf{Method} & \textbf{Dataset Gen.} & \textbf{\#Params} & \textbf{Train time} & \textbf{Infer.} \\
\midrule
MCNet &  --  & 5M & 17min & 7s \\
SSPSR  &  -- & 14M & 34min & 10s \\
HSISR  &  --  & 22M & 18min & 13s \\
MSDformer &  -- & 13M & 43min  & 6s  \\
HyperSIGMA &  -- & 189M & 5.6D + 16min & 12s \\
MSDformer-DL &  55min  & 13M & 8h & 6s \\
MCNet-DL &  55min  & 5M  & 5h & 7s \\
\bottomrule
\end{tabular}
\label{tab_complexity_comparison}
\end{table}

All of the above experiments are conducted on a single A100 GPU under the same hardware configuration (1 GPU, 10 CPU cores, 120 GB RAM). Table~\ref{tab_complexity_comparison} reports, for each method, the dataset generation time (when applicable), the number of parameters, the training time, and the average inference time per hyperspectral image on the Urban dataset with $M = 6$ and a scale factor of $\times 4$. For MSDformer-DL and MCNet-DL, the \textit{Dataset Gen.} column corresponds to the time required to generate 5,000 $(A_{DL,HR}, A_{DL,LR})$ pairs of the same size as the HSI patches for training, whereas the other methods are trained directly on Urban patches. HyperSIGMA is used as a foundation model: its reported training time combines 5.6 days of pre-training with an additional 16 minutes of fine-tuning on Urban. Compared to the other deep-learning SISR methods, MCNet-DL has the smallest number of parameters. This is particularly true compared to the HyperSIGMA foundation model, for which there is a difference of almost 2 orders of magnitude. In terms of train time, our method is the slowest, to the exception of MSDformer-DL. Nonetheless, it is interesting to note that this is mostly due to the fact that our method sees more training samples than the other ones, and therefore requires more training time, and not to the the synthetic data generation itself. To alleviate this computational burden issue, we plan in future work to study the transferability of our trained model to new areas, without retraining or with a light retraining.

\section{Conclusion}

In this work, we introduced a novel approach for the super-resolution of hyperspectral remote sensing images. By designing a framework based entirely on synthetic data, our method operates in an unsupervised manner, thus addressing the challenges related to limited datasets and the absence of ground truth. Significant efforts were dedicated to enhancing the realism of synthetic data, thereby improving both the performance and robustness of network training. Experimental results on three widely used hyperspectral datasets validated the effectiveness of our approach across various down-sampling scales, with performance metrics confirming the model ability to preserve both spatial structures and spectral fidelity.

Moreover, the proposed framework offers considerable flexibility, both in the generation of synthetic data and in the choice of super-resolution networks, allowing for a wide range of adaptations in network design, dataset creation, and their combinations. Nevertheless, the method relies on prior knowledge of the sensor’s PSF to remain effective. One perspective could involve integrating PSF estimation within the framework. Another promising perspective would be to design synthetic abundance datasets with sufficient generality to enable super-resolution of remote sensing abundance independently of sensor-specific characteristics and observed zones.

\section*{Acknowledgments}
\noindent The work was partially supported by Agence de l’Innovation de Défense – AID - via Centre Interdisciplinaire d’Etudes pour la Défense et la Sécurité – CIEDS - (project 2023 - ALIA).

\bibliographystyle{IEEEtran}
\bibliography{biblio}


\begin{IEEEbiography}[{\includegraphics[width=1in,height=1.25in,clip,keepaspectratio]{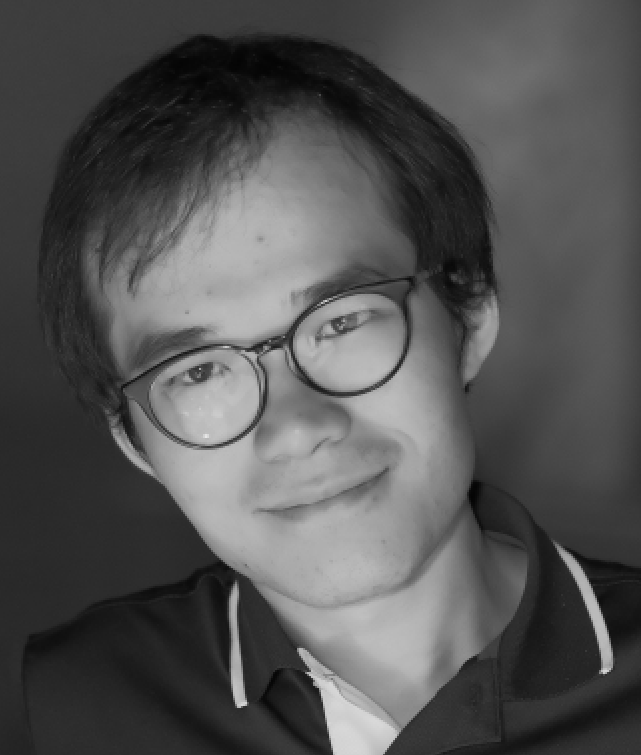}}]{Xinxin Xu} received the engineering degree from Institut d'Optique Graduate School, France, in 2023. He is currently pursuing the Ph.D. degree with LTCI, Télécom Paris, Institut Polytechnique de Paris, Palaiseau, France, under the supervision of Yann Gousseau, Christophe Kervazo and Saïd Ladjal. 

His research focuses on hyperspectral image super-resolution, with applications in remote sensing.
\end{IEEEbiography}

\begin{IEEEbiography}[{\includegraphics[width=1in,height=1.25in,clip,keepaspectratio]{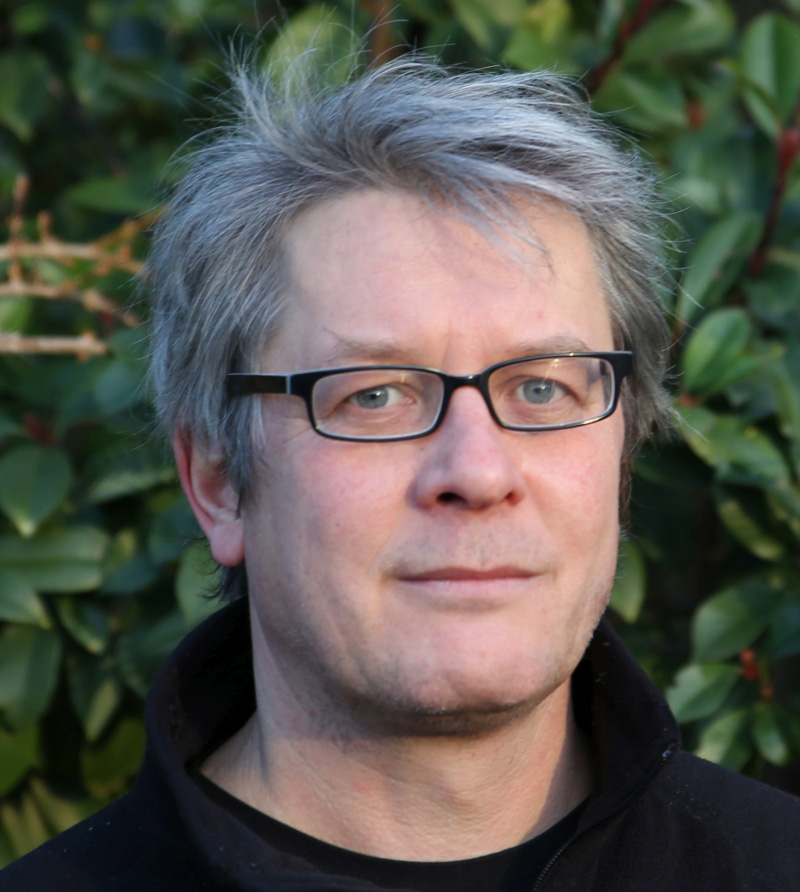}}]{Yann Gousseau} received the B.Eng. degree from the École Centrale de Paris, Châtenay-Malabry, France, the Part III of the Mathematical Tripos degree from the University of Cambridge, Cambridge, U.K., in 1995, and the Ph.D. degree in applied mathematics from the University of Paris-Dauphine, Paris, France, in 2000.

He is currently a Professor with Télécom Paris, Palaiseau, France. His research interests include the mathematical modeling of natural images and textures, generative models, computer vision, image, and video processing.
\end{IEEEbiography}

\begin{IEEEbiography}[{\includegraphics[width=1in,height=1.25in,clip,keepaspectratio]{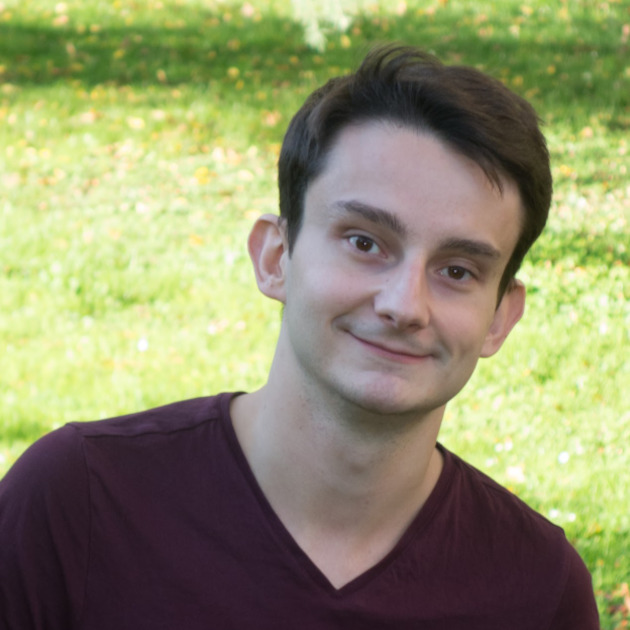}}]{Christophe Kervazo}
 received the engineering degree from Supélec, France, in 2015, as well as the Master of Sciences degree from Georgia Institute of Technology, Atlanta, USA. He did is PhD at CEA Saclay, Gif-sur-Yvette, France.
 
 He is currently associate professor at Télécom Paris, Palaiseau, France, where he works on imaging applications. His research focuses on interpretable deep learning, both from the neural networks architecture point of view and the reliability of their outputs. His applications include remote sensing and medical imaging.
 \end{IEEEbiography}

 \begin{IEEEbiography} [{\includegraphics[width=1in,height=1.25in,clip,keepaspectratio]{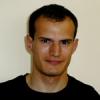}}]{Saïd Ladjal}
 received a diplôme de magister from École normale supérieure in 2000 including a Masters degree in computer science, engineering degree from Télécom Paris in 2002 and Ph.D degree in applied mathematics from École normale supérieure de Cachan in 2005.

 He is currently a Professor with Télécom Paris. His research interests are on mathematical modeling for images and computational photography with applications to remote sensing, general image restoration and medical imaging. 
 \end{IEEEbiography} 

\end{document}